\documentclass[10pt,conference]{IEEEtran}
\IEEEoverridecommandlockouts
\usepackage{cite}
\usepackage{amsmath,amssymb,amsfonts}
\usepackage{algorithmic}
\usepackage{graphicx}
\usepackage{textcomp}
\usepackage{xcolor}

\definecolor{codeColor}{RGB}{139,26,26}

\newcommand{\ali}[1]{\textcolor{blue}{{\it}}}
\newcommand{\mehdi}[1]{\textcolor{red}{{\it}}}
\newcommand{\code}[1]{\textcolor{codeColor}{\textsf{#1}}}

\newcommand{\figref}[1]{Figure~\ref{#1}}

\def\BibTeX{{\rm B\kern-.05em{\sc i\kern-.025em b}\kern-.08em
    T\kern-.1667em\lower.7ex\hbox{E}\kern-.125emX}}
\begin{document}

\title{A Program Synthesis Approach for Adding Architectural Tactics to An Existing Code Base}

\author{\IEEEauthorblockN{Ali Shokri}
\IEEEauthorblockA{\textit{dept. of Software Engineering} \\
\textit{Rochester Institute of Technology}\\
Rochester, NY, USA \\
as8308@rit.edu}
}

\maketitle

\begin{abstract}
%Software designers and developers more increasingly incorporate third-party APIs in their programs to address important architectural tactics. There has been many research conducted on learning the correct way of using APIs in a program for correct implementation of a specific usecase or tactic. These API usage models then, could be used for automatically synthesizing a program  
Automatically constructing a program based on given specifications has been studied for decades. Despite the advances in the field of Program Synthesis, the current approaches still synthesize a block of code snippet and leave the task of reusing it in an existing code base to program developers. 
Due to its program-wide effects, synthesizing an architectural tactic and reusing it in a program is even more challenging. Architectural tactics need to be synthesized based on the context of different locations of the program, broken down to smaller pieces, and added to corresponding locations in the code. Moreover, each piece needs to establish correct data- and control-dependencies to its surrounding environment as well as to the other synthesized pieces. This is an error-prone and challenging task, especially for novice program developers.   

%However, as program-wide tactics, architectural tactics need to be synthesized based on the context of different locations of the program, broken down and each piece be added to the corresponding location in the code. Moreover, each piece needs to establish correct data- and control-dependencies to its surrounding environment as well as other synthesized pieces. This is an error-prone and challenging task, especially for novice program developers. 

%Breaking down the code snippet to meaningful pieces, adding each piece to the correct location in the code, and establishing correct data- and control-dependencies between each piece and its surrounding environment as well as other pieces is an error-prone and challenging task, especially for novice program developers. 

In this paper, we introduce a novel program synthesis approach that synthesizes architectural tactics and adds them to an existing code base.   
\end{abstract}


\section{Background and Research Problem}
Program synthesis aims to automatically construct a program in an underling programming language which satisfies a set of program specifications \cite{gulwani2017program}. Program developers can benefit from the synthesized code either by directly reusing it in their programming tasks, or by learning from it.
%These specifications could be provided in a variety of ways, from a few samples of input/output pairs that represent the expected behavior of the program \cite{X}, to more (semi) formal specifications explicitly described by the programmer \cite{X}, to a mixture of these approaches \cite{X}. 
%While many research focus on synthesizing programs in a Domain Specific Languages (DSL) \cite{X}, there are some work that tackle the problem of program synthesis in a Generic Purpose Language (GPL) \cite{X}. The later type of work aims to assist a wider range of programmers with generating code snippets for different programming tasks. 
Studies \cite{rehman2018roles, santos2017understanding} show that one important programming challenge that novice programmers deal with is to implement architectural tactics, as a powerful means of addressing important quality attributes of a software, in their programs. For instance, adding \textit{\textbf{authentication}} and \textit{\textbf{authorization}} security tactics to a program under-development is a nontrivial and error-prone task for such programmers. These tactics are mostly implemented by incorporating APIs of third-party libraries \cite{cervantes2012principled}. For example, Java Authentication and Authorization Security Services (JAAS) is a popular Java-based framework that provides APIs for adding such tactics to a program. Component-based program synthesis \cite{jha2010oracle}, which aims to create a code snippet only from a list of given components (e.g., APIs), could be an answer to this need. 
In the past recent years, there are some work that specifically focus on synthesizing a program using APIs \cite{feng2017component, yang2018edsynth, shi2019frangel, guo2019program, liu2020prosy, liu2020much}. However, despite the advances in API-based program synthesis, novice programmers still are not able to easily incorporate these approaches in architectural tactic implementation tasks. There are two main barriers in this regard. First, programmers need to express the specification of architectural tactics to be synthesized. Nonetheless, novice programmers have no insight into this type of complicated specification. Second, the synthesized code might need to be broken in smaller pieces, placed in different locations of the under-development program and used in an inter-procedural manner. For instance, in case of authentication, the programmer might need to add the \textit{initialization} process in method \code{m1()} of class \code{C1}, the process of \textit{logging in} in method \code{m2()} of class \code{C2}, and the process of \textit{logging out} in method \code{m3()} of class \code{C3}. Then, calls all the three methods in method \code{m4()} of class \code{C4} sequentially. The current state-of-the-art approaches do not address these needs.

In this paper, we propose a novel approach for synthesizing architectural tactics in a given program. More specifically, we define our research problem as: \textit{Given a compilable program, i.e., syntactically correct program, we aim to automatically synthesize and add architectural tactics to that program such that the final program is syntactically and semantically (w.r.t. APIs) correct.} We build upon our previous work \cite{shokri2021_icpc, shokri2021_icsa} on creating Framework API Specification model which is a comprehensive probabilistic representation of correct ways of using APIs of a specific framework for implementing architectural tactics.

\begin{figure*}[t]
    \centering
    \includegraphics[width=\textwidth]{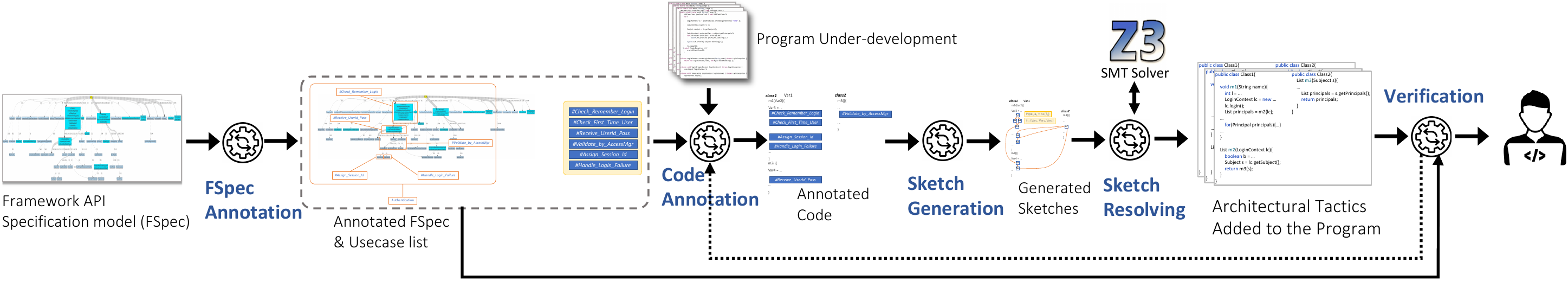}
        \centering
    \caption{An overview of the proposed approach}
    \label{fig:ipps}
\end{figure*}

\section{Related Work}
%\ali{To be completed}
In recent years, a number of API-based program synthesis approaches have been developed. SyPet \cite{feng2017component} synthesizes a block of code based on a given list of APIs, the signature (i.e., type of input and output) of the expected code snippet, as well as some test cases. EdSynth \cite{yang2018edsynth} and FrAngle \cite{shi2019frangel} are able to construct a short code snippet with more control structures compared to SyPet. TYGAR \cite{guo2019program} aims to improve the performance and scalability of previous API-based synthesizers. Although these approaches use APIs as building blocks, they do not use some precious information about APIs, i.e., API usage patterns, in the synthesis process. In some of the follow-up works \cite{liu2019accelerating, liu2020prosy, liu2020much}, researchers incorporated API usage patterns in the synthesis task which resulted in improvement of the performance and accuracy of the synthesizer. Despite the capabilities of all these approaches, they suffer from two shortcomings which makes them incapable of being used in architectural tactic synthesis task. First, the need for providing specifications of the to-be-synthesized code, and second, the lack of ability of synthesizing a discrete (and yet related) code snippets in different locations of the program. In this paper, we introduce an approach that addresses these concerns.

\section{Approach}
\figref{fig:ipps} provides an overview of our approach which consists of the following five main steps. 

\subsection{FSpec Annotation}
We leverage the Framework API Specification model (FSpec) \cite{shokri2021_icsa} that only includes correct API usages for implementing architectural tactics in our program synthesis approach. In that regard, we create a hierarchical annotation of API usages in the FSpec. Each annotation represents a meaningful usage of APIs to accomplish a (sub-) task. Hence, one would expect to see a couple of APIs together annotated as \textit{\#Authentication}, while there are  more fine-grained annotations such as \textit{\#Initialization}, \textit{\#Logging\_in}, and \textit{\#Logging\_out} inside the \textit{\#Authentication}. 
%At the moment, this process is performed manually. 
We use method naming suggestion tools \cite{liu2019learning, zhang2016towards} for the purpose of automatic FSpec annotation. 
%\ali{What is the API grouping mechanism then?}. 

\subsection{Code Annotation}
Next, we find a mapping between each of the selected FSpec annotations and possible locations in the program that the annotation can be added to. This mapping basically shows that \textit{\textbf{what}} sub-FSpec part should be synthesized, and \textit{\textbf{where}} in the program should it be placed. 
%Currently, we ask the user to manually specify the location in the code in which the provided annotations should be added to.
To make this process automatic, we  enhance FSpec mining process with context annotation attachments. More specifically, while creating the FSpec, we attach information about the method name, class name, available variable types, control- and data-flow graph of surrounding environment of API to the FSpec. This enables us to identify and rank the candidate locations for each annotation in the under-development program.

%The \textit{what} part guides us towards the sketch to be generated, and the \textit{where} part would provide the constraints based on the location in the code. For instance, if API \code{$A_1$} takes a \code{String} as an input, we perform an inter-procedural static analysis to find out what variables of type \code{String} are visible at that location which are already initialized and ready to use.  

\subsection{Sketch Generation}
Each annotation consists of a set of APIs and their dependencies (i.e., control and data dependencies) represented as a graph in which the nodes are APIs and edges are their dependencies. We translate this graph to a code snippet that is only composed of those APIs. This code snippet is in the format of Static Single Assignment (SSA). If there is a data dependency between API \code{$A_1$} and \code{$A_2$} (i.e., API \code{$A_2$} uses the data generated by \code{$A_1$}), the output of \code{$A_1$} is an input to \code{$A_2$} in the generated code snippet. However, there are still two  dependencies remained to be added to this code snippet \textbf{(i)} dependencies between the code snippet and variables defined earlier in the program that are visible to the code snippet, and \textbf{(ii)} dependencies between this code snippet and the other generated code snippets, i.e., dependencies between annotations. To address these dependencies, we add holes to those arguments of the APIs in the code snippet that are not already filled. The output of this step would be a \textbf{\textit{sketch}} of the final version of the to-be synthesized code snippet. 

\subsection{Sketch Resolving}
In the next step, we translate the sketched code snippet to an SMT problem and use an off the shelf SMT solver, Z3, to find a corresponding solution. More specifically, we statically analyze the program and find all the visible variables that are already initialized and have the same type as each hole. \ali{More constraints like an oracle?} Then, we translate back the solved SMT problem to our actual sketched code snippet problem and replace holes with the found variable names.   

\subsection{Verification}
Finally, to make sure that the synthesized code snippets in the program are correctly implementing the expected architectural tactic, we perform a static analysis over the program and create its inter-procedural API usage model with respect to the framework of interest. We use our previously developed technique, ArCode \cite{shokri2021_icpc}, for this purpose. In case that no deviation from the correct tactic implementation is detected, we consider the overall process as a successful program synthesis. Otherwise, a feedback would be sent to \textit{Code Annotation} process to learn from this failure and use it for the next round of finding candidate locations for annotations. 

\section{Conclusion and Future Work}
In this research, we introduced a novel program synthesis-based approach for adding architectural tactics to an existing code base. This is an ongoing research in which we are currently able to generate code snippets for annotations.  
%This approach is able to break down the to-be-synthesized code into smaller pieces of sketches,  add each part to a candidate location in the code, resolve all the sketches, and implement the desired architectural tactic in the program.   

The output of this research can enable many software development tools (e.g., IDEs) to support programmers with automation of architectural tactic  recommendation and implementation. Moreover, realization of this approach as a tool could serve in educational purposes such as teaching software architecture to program developers. 

\bibliographystyle{IEEEtran}
\bibliography{IEEEabrv,Bibliography}

\end{document}